\documentclass[10pt,journal]{IEEEtran}
\newtheorem{theorem}{Theorem}

\newtheorem{lemma}{Lemma}
\newtheorem{definition}{Definition}
\newtheorem{remark}{Remark}

\newtheorem{assumption}{Assumption}

\usepackage{graphicx}
\usepackage{subfigure}
\usepackage{array}
\usepackage{amsmath}
\usepackage{amssymb}
\usepackage{cite}
\usepackage{enumerate} 
\usepackage{float}
\usepackage{amsfonts} 
\usepackage{epsfig} 
\usepackage{epstopdf}
\usepackage[dvipsnames]{xcolor}
\usepackage{tikz}\newcommand*{\circled}[1]{\lower.7ex\hbox{\tikz\draw (0pt, 0pt)%
 		circle (.4em) node {\makebox[1em][c]{\small #1}};}}
\usepackage{color}
\usepackage[scaled]{helvet}

\begin{document}

\title{Adaptive Control with Global Exponential Stability  for
	Parameter-Varying Nonlinear Systems under
	Unknown Control Gains}
\author{Hefu Ye, Haijia Wu, Kai Zhao,  and Yongduan Song$^{*}$, \IEEEmembership{Fellow, IEEE}
	\thanks{This work was supported in part by the Graduate Research and Innovation Foundation of Chongqing, China under grant CYB22065, and in part  by the National Natural Science Foundation of China under grant (No.61991400, No.61991403, No.61860206008, No.61933012, and No.61833013). Corresponding author is Yongduan Song.}
	\thanks{H. Ye  and Y.   Song are with the Chongqing Key Laboratory of Autonomous Systems, Institute of Artificial Intelligence, School of Automation, Chongqing University, 400044, China.  (e-mail:    yehefu@cqu.edu.cn; ydsong@cqu.edu.cn).}
	\thanks{H. Wu is with the Army Logistics Academy, Chongqing, 401331, China (e-mail: 	deepbodhi@icloud.com).}
	\thanks{K. Zhao is with the Department of Electrical and Computer Engineering, National University of Singapore, 119077, Singapore (e-mail: 		zhaokai@cqu.edu.cn).}
}

\maketitle     
\begin{abstract} 	
	It is nontrivial to achieve exponential stability even for time-invariant nonlinear systems with matched uncertainties and persistent excitation (PE) condition. In this paper, without the need for  PE condition, we address the problem of global exponential stabilization of strict-feedback systems with mismatched uncertainties and unknown yet time-varying control gains. The resultant control, embedded with time-varying feedback gains, is capable of ensuring global exponential stability of parametric-strict-feedback systems in the absence of persistence of excitation. By using the enhanced Nussbaum function, the previous results are extended to more general nonlinear systems	where the sign and magnitude of the time-varying control gain are unknown. In particular,  the argument of the Nussbaum function is guaranteed to be always positive with the aid of nonlinear damping design, which is critical to perform a straightforward technical analysis of the boundedness of the Nussbaum function. Finally, the global exponential stability of parameter-varying strict-feedback systems, the boundedness of the control input and the update rate, and the asymptotic constancy of the parameter estimate are established. Numerical simulations are carried out to verify the effectiveness and benefits of the proposed methods. 
\end{abstract}


\section{Introduction}
\label{sec:introduction}

\IEEEPARstart{T}{he} past few decades have witnessed extensive developments and applications of nonlinear adaptive control for dynamic systems with unknown parameters \cite{Isidori,Praly,Taylor,Kanelakopoulos,krstic,WenChangyun,Ioannou-PE,zhuanghuaguang}. In the early works on adaptive control \cite{Isidori,Praly,Taylor}, certain restrictions, such as matching condition, extended matching condition and growth conditions on system nonlinearities, are normally imposed. Adaptive backstepping technology \cite{Kanelakopoulos,krstic}, on the other hand, has  completely removed these restrictive conditions, thus motivating considerable amount of studies on adaptive control of various systems (see, for instance, \cite{WenChangyun,Ioannou-PE}).  It is noted that most existing works focus primarily on systems with constant parameters and known control directions.

To address the adaptive estimation of time-varying parameters, the pioneering work \cite{Goodwin-PE} proposes a method to exponentially stabilize  linear time-varying systems in virtue of persistent excitation (PE) condition. Subsequently, such condition is removed in \cite{Goodwin-nonPE}  and the linear model is extended to robot system in \cite{song1992tac}.  Thereafter, the output feedback scheme is studied in \cite{Marino-TAC}, where the so-called  projection operation is exploited  to guarantee the boundedness of slow time-varying parameter estimate.  For more general systems, such as strict-feedback systems, the soften sign function is introduced to cope with unknown parameters in \cite{Ioannou-PE} and \cite{WenChangyun}. More recently, an elegant adaptive scheme based on  ``congelation of variables" method is proposed in \cite{chenkaiwen,chenkaiwenACC,chenkaiwenIFAC}  to asymptotically stabilize parametric strict-feedback systems with fast time-varying parameters,  opening a new venue for developing certainty equivalence controller for nonlinear time-varying systems.  This method is also used to address the formation tracking of  multi-agent systems in \cite{ChenYY} and \cite{ChenYY2}.

Interest in adaptive control of nonlinear systems with unknown control direction is stimulated by the development of Nussbaum functions \cite{Nussbaum} and the corresponding lemmas for stability analysis \cite{yexudong-auto,yexudong-tac}. The idea behind such control is that by specifying a controller with a Nussbaum gain, so that the controller degrades the system performance in the period with a wrong direction, but rewards the system with quicker movement to the desired state with a higher gain when the sign alternates to a correct direction in the subsequent period. In \cite{SSGe2004,huachangchun,Liuyanjun}, based upon this idea, the adaptive neural control, adaptive fuzzy control and adaptive guaranteed performance control are proposed for nonlinear systems without \textit{a priori} knowledge of the control direction, but these results need  the precise information of the control gain magnitude.  When the control coefficient is unknown both in sign and magnitude,  some robust and adaptive results (\textit{e.g.,} \cite{SSGe2003,Huang2018}) achieving asymptotic stability of the closed-loop system are established through a particular Nussbaum function $\exp({x^2})\cos(\pi x/2)$.  Recently, the results in \cite{2015-chen-stabilization} and \cite{chenzhiyongauto} invent  a class of enhanced Nussbaum functions to deal with time-varying and/or multi-variable unknown control coefficients.  The introduction of the Nussbaum functions not only presents the likelihood of realization of the adaptive scheme but also enables some practical applications. Hence, it is meaningful and necessary to address the issue of adaptive control for nonlinear systems with time-varying parameters and unknown control coefficient (including control direction and control gain magnitude).

Motivated by the above analysis, here in this work we develop an adaptive control method to guarantee global exponential stability of parameter-varying strict-feedback systems with unknown control direction. The global exponential stability of nonlinear systems is under explored even if the systems do not involve time-varying parameters and/or unknown control coefficient. The original works that employ acceleration control (with exponential convergence) for stabilizing strict-feedback systems are \cite{Zhao}, where a time-varying scaling is used to accelerate the original system dynamics; after stabilizing the accelerated system, the convergence rate of the original system can be assigned by incorporating a suitable time-varying function into the control scheme. Finally, the developed adaptive schemes are verified on the wing-rock model.
The contributions of this article are three-fold:
\begin{itemize} 
	\item Different from the asymptotic results in \cite{chenkaiwen,chenkaiwenACC,chenkaiwenIFAC} relying on the \textit{a priori} knowledge of the control direction, our algorithm achieves exponential stability for parameter-varying nonlinear systems and is effective for unknown control direction;   
	\item Exponential convergence is normally realized  at the rather restrictive  PE condition, here we achieve exponential convergence without PE. In addition, the proposed method is capable of dealing with both time-varying parameters and unknown control coefficient, covering the existing one \cite{Zhao}  on exponential stabilization  as a special case; 
	\item Although there exist  prescribed performance control results capable of ensuring  semi-global exponential convergence,  those methods do not guarantee zero-error regulation for nonlinear systems\cite{Benchlioulis,zhangcyber-global,Wang2017,wangqing,dongxiwang,wang-smc-2016},  the proposed control method guarantees global exponential regulation with zero steady-state error  (more favorable performance) for nonlinear systems with unknown time-varying parameters and unknown control coefficients.
\end{itemize}


The remainder of this paper is organized as follows. In Section \ref{section2}, we state the necessary mathematical preliminaries and a key lemma for our main results.  We introduce some scalar examples in Section \ref{section3}, to present the main design ideas before turning to general designs. In Section \ref{section4A}, an adaptive control scheme is proposed to exponentially stabilize  the  time-varying nonlinear system with known control direction; in Section \ref{section4B},  we extend this scheme to  time-varying parametric strict-feedback systems with unknown control coefficient. In Section \ref{section5}, a comparative simulation is considered to illustrate the main results. Finally, conclusions and perspectives are given in Section \ref{section6}. 

 \textit{Notations:} 
$\mathbb{R}$ is the field of reals, $\mathbb{R}^+=\{a\in\mathbb{R}: a >0\}$, $\mathbb{R}^n$ denotes the $n$-dimensional Euclidean space, and	$\mathbb{R}^{n\times m}$ is the set of $ n \times m$ real matrices. $\underline{x}_i=[x_1,\cdots,x_i]^{\top}\in\mathbb{R}^i$ denotes a vector. $\Gamma \succ 0$  means that the symmetric matrix $\Gamma$ with suitable dimensions is positive definite.  $|W|_{_{\mathsf{F}}}=\sqrt{\sum_{i=1}^{n}\sum_{j=1}^{q}({W}_{ij})^2}$ denotes the Frobenius norm of matrix $W$. $f\in\mathcal{C}^{\infty}$  denotes a function $f$ has continuous derivatives of order $\infty$, and $f\in\mathcal{L}_\infty$ denotes a function $f$ is bounded. In addition, $\mathcal{L}_1$ and $\mathcal{L}_2$ denote the classes of Lebesgue-integrable functions.

\section{preliminaries}\label{section2}
\subsection{System Description \& Assumptions}
Consider the following single-input single-output nonlinear systems with time-varying parameters\cite{chenkaiwen}:
\begin{equation}\label{system}
\left\{\begin{array}{ll}
\dot{x}_1=\phi_1^{\top}(  {x}_1)\theta(t)+x_{2}\\
~~~\vdots\\
\dot{x}_i=\phi_i^{\top}(\underline {x}_i)\theta(t)+x_{i+1}\\
~~~\vdots\\
\dot{x}_n=\phi_n^{\top}(\underline{x}_n)\theta(t)+b(t)u
\end{array}\right.
\end{equation}
where $x=[x_1,\cdots,x_n]^{\top}\in\mathbb{R}^n$ and $u\in\mathbb{R}$ are the state vector and the control, respectively. The regressors $\phi_i:\mathbb{R}^i\rightarrow\mathbb{R}^q,~i=1,\cdots,n,$ are smooth mappings and satisfy $\phi_i(0)=0.$ By the mean value theorem, there exist continuous matrix-valued functions $\Phi_i(\underline{x}_i)\in\mathbb{R}^{q\times i}$ such that $\phi_i(\underline{x}_i)=\Phi_i(\underline{x}_i)\underline{x}_i$.
The system parameters  $\theta(t)\in\mathbb{R}^q$ and $b(t)\in\mathbb{R}$ are unknown and time-varying parameters and satisfy the following assumptions:\footnote{As $b(t)=\operatorname{sgn}(b(t))|b(t)|$, throughout this paper,  $b(t)$ is referred to as the control coefficient,  $\operatorname{sgn}(b(t))$ the control direction,   and $|b(t)|$ the control gain magnitude.}

\begin{assumption}[{[\citen{chenkaiwen}]}]\label{assumption1}
	The parameter $\theta(t)$ is piecewise continuous and $\theta(t)\in \Theta_0,$   $\forall t\geq 0$, where $\Theta_0$ is a completely unknown compact set.   The ``radius" of $\Theta_0$, denoted by $\delta_{\Delta_{\theta}}$, is assumed to be known, while $\Theta_0$ can be unknown.
\end{assumption}

\begin{assumption}[{[\citen{chenkaiwen}]}]\label{assumption2}
	The control direction is known and does not change. We assume that  $b(t)$  is unknown but bounded away from zero in the sense that there exists an unknown constant $\ell_b$, such that   $0< |\ell_b| \leq  |b(t)| $, for all $t\geq 0$.
\end{assumption}

\begin{assumption}[{[\citen{2015-chen-stabilization} Section 6.3]}]\label{assumption3}
	The time-varying control coefficient $b(t)\notin 0$ take values in an unknown compact set $\Omega_b $.  The control direction is unknown and does not change.
\end{assumption}
 
\begin{remark}
In our later control development, we present two set of control schemes:  the first one (Theorem 1) is based on Assumption 2,  in which the control gain magnitude  is allowed to be unknown and time-varying, and the second one (Theorem 2) is built upon Assumption 3, in which both the control direction and the control gain magnitude are  allowed to be unknown.
\end{remark}

\subsubsection*{Control Objective} The control objective in this paper is to design  adaptive control schemes for system (\ref{system}) to achieve globally exponential stabilization.   

This problem has received a lot of attentions. But until recently, the contributions were only achieving that asymptotic stabilization and/or bounded stabilization as in \cite{chenkaiwen,Huang2018}. The exponentially stable results can only be obtained under the  persistence of excitation or under the assumption that $\theta$ is  constant and $b(t)=1$  as in \cite{Zhao,Goodwin-PE}.

\subsection{Enhanced Nussbaum function \& Corresponding Lemma}
\begin{definition}[{[\citen{chenzhiyongauto}]}]
	A $\mathcal{C}^{\infty}$ function $\mathcal{N}(\xi):[0, \infty)$ $\mapsto(-\infty, \infty)$ is called an enhanced Nussbaum function if it satisfies
	\begin{equation*}
	\begin{aligned}
	\lim _{\xi \rightarrow \infty} \frac{1}{\xi}{\int_{0}^{\xi} \mathcal{N}^{+}(\tau) d \tau} &=\infty,~~~ \limsup _{\xi \rightarrow \infty} \frac{\int_{0}^{\xi} \mathcal{N}^{+}(\tau) d \tau}{\int_{0}^{\xi} \mathcal{N}^{-}(\tau) d \tau}=\infty, \\
	\lim _{\xi \rightarrow \infty}   \frac{1}{\xi}{\int_{0}^{\xi} \mathcal{N}^{-}(\tau) d \tau}  &=\infty, ~~~\limsup _{\xi \rightarrow \infty} \frac{\int_{0}^{\xi} \mathcal{N}^{-}(\tau) d \tau}{\int_{0}^{\xi} \mathcal{N}^{+}(\tau) d \tau}=\infty.
	\end{aligned}
	\end{equation*}
	where $\mathcal{N}^+$ and $\mathcal{N}^-$ denote the  positive and negative truncated functions of $\mathcal{N},$ respectively, \textit{i.e.,} $\mathcal{N}^{+}(\xi)=\max \{0, \mathcal{N}(\xi)\},~ \mathcal{N}^{-}(\xi)=\max \{0,-\mathcal{N}(\xi)\}$. A legal fraction (with non-zero denominator) assumption is implicitly made in the above  definition  which excludes the trivial function $\mathcal{N}(\xi) \equiv 0$.
\end{definition}


\begin{lemma}[{[\citen{chenzhiyongauto}]}]\label{lemma1}
	Consider two  $\mathcal{C}^{\infty}$ functions $ {V}(t)$ : $[0, \infty) \mapsto \mathbb{R}^{+}, \mathcal{N}(t): [0, \infty) \mapsto \mathbb{R} ^{+}$. Let $b(t):[0, \infty) \mapsto[\underline{b}, \bar{b}]$ for two constants $\underline{b}$ and $\bar{b}$ satisfying $\underline{b} \bar{b}>0$. If
	\begin{equation*}
	\begin{aligned}
	&\dot{V}(t) \leq(b(t) \mathcal{N}(\xi)+1) \dot{\xi}(t) \\
	&\dot{\xi}(t) \geq 0, \quad \forall t \geq 0
	\end{aligned}
	\end{equation*}
	for an enhanced Nussbaum function  $\mathcal{N}$, then $V(t)$ and $\xi(t)$ are bounded over $[0, \infty)$.
\end{lemma}

\section{The basic design ideas}\label{section3}
This section introduces the basic design ideas  through three simple scalar systems. 
\subsection{Exponential regulation for systems with time-invariant parameters}\label{sectionA}
 To obtain a better understanding of how time-varying scaling may be used to achieve exponential regulation, we consider 
\begin{equation}\label{example1}
\dot x= u+ax^2
\end{equation}
where $x\in\mathbb{R}$ and $u\in\mathbb{R}$ are the state and the control, respectively, and $a$ is an unknown constant. Let $$s=e^{\lambda t}x\triangleq \mu(t)x$$ with $\lambda>0$ being the acceleration constant,  then
\begin{equation}\label{s3.1}
\dot s= \mu(\lambda x +u+\hat{a}x^2)+\mu(a-\hat{a})x^2.
\end{equation}
Choose the Lyapunov function as
\begin{equation}\label{V3.1}
V= \frac{1}{2}s^2+\frac{1}{2\gamma_{a}}(a-\hat{a})^2
\end{equation}
where $\gamma_{a}>0$ and $\hat{a}$ is the estimation of $a$, then
\begin{equation}\label{dV3.1}
\dot V=  \mu s   \left(\lambda x+u+\hat{a}x^2 \right)+\frac{1}{\gamma_{a}} (a-\hat{a})\left(\gamma_{a}\mu s x^2-\dot{\hat{a}}\right).
\end{equation}
Design the update law and control law as
\begin{equation}\label{u3.1}
\begin{aligned}
&\dot{\hat{a}}=\gamma_{a}\mu s x^2\\
&u=-(k+\lambda)x-\hat{a}x^2.
\end{aligned}
\end{equation}
where $k>0$. Substituting (\ref{u3.1}) into (\ref{dV3.1}), we get
\begin{equation}\label{dV3.1.1}
\dot V= -ks^2\leq 0.
\end{equation}
It follows from (\ref{dV3.1.1}) that $V\in \mathcal{L}_{\infty}$ and therefore the boundedness of $s$ and $\hat{a}$ is guaranteed. In fact, the boundedness and the exponential convergence of $x$ are established simultaneously since $x=e^{-\lambda t}s$, which further implies the boundedness of $u$. To show the asymptotic constancy of $\hat{a}(t)$, note that $\dot{\hat{a}}=\gamma_{a} \mu s x^2=\gamma_{a} x s^2$ and $x(t)$ is a bounded function, then there exists a number $L$ such that $|\dot{\hat{a}}|\leq  L s^2$. It is seen from  (\ref{dV3.1.1}) that $s\in\mathcal{L}_{2} $ and hence $\dot{\hat{a}}\in\mathcal{L}_1$, then   by using the argument of Theorem 3.1 in \cite{Krstic1996}, it is concluded that $\hat{a}$ has a limit as $t\rightarrow\infty$.

\subsection{Exponential regulation for systems with time-varying parameters}
Here we show how to extend the aforementioned method to exponentially stabilize the systems with time-varying parameters. Consider
\begin{equation}\label{example2}
\dot x= u+a(t)x^2
\end{equation}
where $x\in\mathbb{R}$ and $u\in\mathbb{R}$ are the state and the control, respectively, and $a(t)$ satisfies Assumption 1. Similar to the state scaling by a $t$-dependent function $\mu(t)$  as in Section-\ref{section3}-A, we define $s=\mu(t)x$, then the dynamics of $s$ becomes
\begin{equation}\label{ds} 
\begin{aligned}
	\dot s=&\mu (u+a(t)x^2)+ \mu \lambda x\\
	=& \mu (\lambda x +\hat{a}x^2 +u )+ \mu(a(t)-\ell_a)x^2+\mu(\ell_a-\hat{a})x^2,
\end{aligned} 
\end{equation}
where $\ell_a$ is an unknown constant, $\gamma_{a}>0$ and $\hat{a}$ is the estimation of $a$.
Choose the Lyapunov function as
\begin{equation}\label{V3.2}
V= \frac{1}{2}s^2+\frac{1}{2\gamma_{a}}(\ell_a-\hat{a})^2
\end{equation}
then
\begin{equation}\label{dV3.2}
\begin{aligned}
\dot V=&\mu s\left(\lambda x+\hat{a}x^2+u\right)+\mu(a(t)-\ell_a) s x^2\\
&+\frac{1}{\gamma_{a}}(\ell_a-\hat{a})\left(\gamma_{a}\mu s x^2-\dot{\hat{a}}\right).
\end{aligned}
\end{equation}
Denote $a(t)-\ell_a$ by $\Delta_{a}$.  By choosing $u$ and $\dot{\hat{a}}$ as  
\begin{equation}\label{u3.2}
\begin{aligned}
&\dot{\hat{a}}=\gamma_{a}\mu s x^2\\
&u=-(k+\lambda)x  -\hat{a}x^2+v,
\end{aligned}
\end{equation}
where $k>0$, and $v$ is an auxiliary input, we get 
\begin{equation}\label{dV3.2.1}
\dot V=-ks^2+\mu s v +\mu\Delta_{a}s x^2. 
\end{equation}
Note that the uncertain term $\mu\Delta_{a}s x^2=\Delta_a s^2x$, and satisfies the matching condition, therefore, one can design $v=- {\delta_{\Delta_{a}}} x^3/2- {\delta_{\Delta_{a}}} x/2$, with $\delta_{\Delta_{a}}>|\Delta_{a}|$,  to eliminate the effect of $\mu \Delta_{a}sx^2$. Here $\delta_{\Delta_{a}}$ is assumed to be a known constant, if $\delta_{\Delta_{a}}$ is unknown, it is also easy to develop a classical adaptive law to build an ``estimate" of $\delta_{\Delta_{a}}$. Therefore, the final control input is 
\begin{equation}\label{u}
u=-(k+\lambda)x  -\hat{a}x^2- \frac{\delta_{\Delta_{a}}}{2} x^3 - \frac{\delta_{\Delta_{a}}}{2} x
\end{equation}
and the derivative of $V$ satisfies  $\dot V\leq-ks^2$.  
Therefore we can conclude boundedness of all trajectories of the closed-loop system as well as convergence of $x$ to zero using the same argument as the one used in Section \ref{sectionA}. 

\begin{remark}
 Note that the underlying issue becomes more difficult when the control coefficient $b(t)$ is unknown, even if the direction of control is known. To overcome this technical obstacle, we need to design a separate adaptive law to estimate $\ell_b$ (refer Assumption 2 for the meaning of $\ell_b$) while dealing with the time-varying part of $b(t)-\ell_b$ by deliberately designing a negative feedback gain. This will be shown in Section \ref{section4}. 
\end{remark}

\subsection{Exponential regulation for systems with time-varying parameters and unknown control coefficients}
Now we further consider the scenario that both unknown fast time-varying parameters and unknown time-varying control coefficient are involved. We show how to integrate enhanced Nussbaum function with the congelation of variables method as well as state scaling to design accelerated adaptive control to achieve exponential regulation without the need for PE condition. Consider 
\begin{equation}\label{example3}
\dot x= b(t)u+a(t)x^2
\end{equation}
where $x\in\mathbb{R}$ and $u\in\mathbb{R}$ are the state and the control, respectively,  $a(t)$ satisfies Assumption 1, and $b(t)$ satisfies Assumption 3.  Let $s=\mu(t)x$ with $\mu(t)=e^{\lambda t}$. Now choose $$V=  \frac{1}{2}s^2+\frac{1}{2\gamma_{a}}(\ell_a-\hat{a})^2$$ then 
\begin{equation}\label{dV3.3}
\begin{aligned}
\dot V=&\mu s\left(\lambda x+\hat{a}x^2+b(t)u\right)+\mu(a(t)-\ell_a) s x^2\\
&+\frac{1}{\gamma_{a}}(\ell_a-\hat{a})\left(\gamma_{a}\mu s x^2-\dot{\hat{a}}\right).
\end{aligned}
\end{equation}
Design
\begin{equation}\label{u3.3}
\begin{aligned} 
&u=\mathcal{N}(\xi)\bar u,\\
&\dot{\xi}=\mu s \bar u.
\end{aligned}
\end{equation}
where $\mathcal{N}(\xi)$ is an enhanced Nussbaum function as described in Definition 1.  
Substituting (\ref{u3.3})   into (\ref{dV3.3}), yields
\begin{equation}\label{dV3.3.1}
\begin{aligned}
\dot V=&\mu s \left(\lambda x+\hat{a}x^2-\bar{u}\right) +\mu(a(t)-\ell_a)s x^2\\
&+\frac{1}{\gamma_{a}}(\ell_a-\hat{a})\left(\gamma_{a}\mu s x^2-\dot{\hat{a}}\right)\\
&+\left(b(t)\mathcal{N}(\xi) +1\right)\dot{\xi}  .
\end{aligned}
\end{equation}
Note that although the first two lines of (\ref{dV3.3.1}) and (\ref{dV3.2}) has a similar structure, we cannot design $\bar u$ directly from formula (\ref{u}), because there is a  constraint in Lemma 1, namely it requires that $\dot\xi(t)\geq 0$. To solve this technical obstacle, we design 
\begin{equation}\label{u3.3.1}
\begin{aligned}
&\dot{\hat{a}}=\gamma_{a}\mu s x^2\\
&\bar u= (k+\lambda)x+\kappa(\hat{a},x)x , 
\end{aligned}
\end{equation}
where $k>0$ and $\kappa$ is a positive function, which will be designed below. As a result, (\ref{dV3.3.1}) becomes
\begin{equation}\label{dV3.3.2}
\begin{aligned}
\dot V=&-\kappa(\hat{a},x) s^2 +\mu  \hat{a}sx^2+ \mu \Delta_{a} s x^2\\
&-ks^2+\left(b(t)\mathcal{N}(\xi) +1\right)\dot{\xi} ,
\end{aligned}
\end{equation}
where $\Delta_{a}\triangleq a(t)-\ell_a$. Applying Young's inequality, we have 
\begin{equation*}\label{leq}
\begin{aligned}
\mu  \hat{a}sx^2+ \mu \Delta_{a} s x^2&=  \hat{a}s^2x+   \Delta_{a} s^2 x\\
&\leq \frac{1}{2}\left((\hat{a}x )^2+1\right)  s^2+   \frac{\delta_{\Delta_{a}}}{2}\left(x^2+1\right) s^2   .
\end{aligned}
\end{equation*}
Therefore, we select $\kappa(\hat{a},x)$ as
\begin{equation}
\kappa(\hat{a},x)=\frac{1}{2}\left((\hat{a}x )^2+1\right)  +   \frac{\delta_{\Delta_{a}}}{2}\left(x^2+1\right)  
\end{equation}
with $\delta_{\Delta_{a}}\geq |\Delta_{a}|$ is a known constant.   Now the derivative of $V$ satisfies  
\begin{equation}\label{dV3.3.3}
\begin{aligned}
\dot V\leq-ks^2+\left(b(t)\mathcal{N}(\xi) +1\right)\dot{\xi}\leq   \left(b(t)\mathcal{N}(\xi) +1\right)\dot{\xi}.
\end{aligned}
\end{equation}
Recalling that $\dot{\xi}=\mu s \bar u =(k+\lambda+\kappa(\hat{a},x))s^2\geq 0$, therefore we can conclude that $V(t)$ and $\xi(t)$ are bounded by virtue of Lemma 1.  Then, the boundedness of all trajectories of the
closed-loop system as well as convergence of $x$ to zero can be guaranteed by using the same argument as the one used in Section \ref{sectionA}.

\begin{remark}\label{remark-basic-idea} It is worth noting that the classical adaptive parameter update law and adaptive control law are equivalent to that designed in (\ref{u3.1}) if the acceleration constant $\lambda=0$. This phenomenon  is easy to understand since $\lambda=0$ can be regarded as the dynamics of the original system has not been changed. In addition, the classical adaptive control law for time-invariant parameter is equivalent to the one design in (\ref{u3.2}) for time-varying parameters if $\delta_{\Delta_{\theta}}=0$, which is also  easy to understand because  $\delta_{\Delta_{\theta}}=0$ means that $\ell_a=a(t)$, namely the parameter will not change with time. In terms of technical realization, these findings are very useful for the controller design of higher-order systems, playing an inspiring role in the design of parameter update laws and  control laws. 
\end{remark}

\section{Main Results}\label{section4}
{Motivated by the appealing features about the time-varying scaling design method, we further explore its applicability to  parametric strict-feedback systems with unknown time-varying parameters and unknown control coefficient. The high-order controller design is based upon Backstepping technology \cite{krstic}. For each step $i, (i = 1, \cdots , n),$ define the coordinate transformation as
	\begin{equation}\label{error variables}
	\begin{aligned}
	z_1&= x_1, \\
	z_i&= x_i-\alpha_{i-1}, ~i=2,\cdots,n
	\end{aligned}
	\end{equation}
	the exponential scaling
	\begin{equation}\label{exponential scaling}
	\begin{aligned}
	s_i&=e^{\lambda t}z_i=\mu(t)z_i, ~\lambda>0
	\end{aligned}
	\end{equation}
	the new regressor vectors
	\begin{equation}\label{new regressor vectors}
	w_{i}\big(\underline{x}_{i}, \hat{\theta}\big)=W_i^{\top} \underline{z}_i=\phi_{i}-\sum_{j=1}^{i-1} \frac{\partial \alpha_{i-1}}{\partial x_{j}} \phi_{j},~\alpha_0=0.
	\end{equation}
	\begin{remark}\label{remmark W}
		Once we have $\alpha_i$ is a smooth and bounded function and $\alpha_i(0,\hat{\theta})=0$,  then from the  coordinate transformation between $\underline{z}_i$ and $\underline{x}_i$ we know that $\underline{x}_i = 0\Leftrightarrow \underline{z}_i = 0$. To design low-conservative control algorithm while guaranteeing zero steady-state error, we adopt regression matrices design approach. Specifically, by using the mean value theorem, we can  express $w_i=\phi_i-\sum_{j=1}^{i-1} \frac{\partial \alpha_{i-1}}{\partial x_{j}} \phi_{j}$ as
		\begin{equation} \label{23}
		\begin{aligned}
		w_i=&\Phi_i \underline{x}_i - \sum_{j=1}^{i-1} \frac{\partial \alpha_{i-1}}{\partial x_{j}} \Phi_{j}\underline{x}_j\\
		=& W_{\underline{x}_i}^{\top}\underline{z}_i - \sum_{j=1}^{i-1} \frac{\partial \alpha_{i-1}}{\partial x_{j}}  W_{\underline{x}_j}^{\top}\underline{z}_j=W_i^{\top}\underline{z}_i
		\end{aligned}
		\end{equation}
		where $W_{\underline{x}_i}\in\mathbb{R}^{i\times q}$, $W_{\underline{x}_j}\in\mathbb{R}^{j\times q}$ and $W_{i}\in\mathbb{R}^{i\times q}$ are known smooth and bounded mappings, whose analytic expressions can be derived from the corresponding exact $\Phi_i$ and $\alpha_i$.
	\end{remark}
	
 Define the tuning functions as 
	\begin{equation}\label{taui}
	\tau_{i}\big(\underline{x}_{i}, \hat{\theta},\mu\big)=\tau_{i-1}+e^{\lambda t}w_{i} s_{i}=  \sum_{j=1}^{i} \mu w_{i} s_{i}.
	\end{equation}
	To clearly describe the section's organization, we introduce some auxiliary functions  before giving the main theorems
	\begin{equation} 
\begin{aligned}
		 \zeta_i= \lambda+\frac{1}{2}\left((n+1-i)\delta_{\Delta_{\theta}}+\frac{1}{\epsilon_{\psi}}+\delta_{\Delta_{\theta}}|W_i|_{_{\mathsf{F}}}^2\right),  
		\end{aligned}
	\end{equation}
		\begin{equation} \label{u4.2}
\begin{aligned} \kappa(\underline{x}_n,\hat{\theta},\mu)= k_n+\lambda+\frac{1}{2}\left( \delta_{\Delta_{\theta}}(|W_n|_{_{\mathsf{F}}}^2+1)+\frac{1}{\epsilon_{\psi}}+\epsilon_{\psi}|\bar{\psi}|^{2}  \right),
		\end{aligned}
	\end{equation}
	where $\delta_{\Delta_{\theta}}>0,~\epsilon_{\psi}>0$,  and $\bar\psi\in\mathbb{R}^n$ is a smooth mapping and satisfying $\psi=\bar{\psi}^{\top}\underline{z}_n$, and
	\begin{equation}\label{psi}
	\begin{aligned}
	\psi=&z_{n-1}+w_n^{\top}\hat{\theta}-\sum_{i=1}^{n-1}\frac{\partial \alpha_{n-1}}{\partial x_i}x_{i+1}-\frac{\partial \alpha_{n-1}}{\partial \hat{\theta}}\tau_n\\
	&-\frac{\partial \alpha_{n-1}}{\partial \mu}\lambda\mu-\sum_{i=2}^{n-1}\frac{\partial \alpha_{i-1}}{\partial \hat{\theta}}\Gamma \mu s_{i}w_n,
	\end{aligned}
	\end{equation}
	where $\Gamma=\Gamma^{\top}\succ 0$ is a matrix.

	\subsection{Adaptive control for parameter-varying nonlinear systems with known control direction}\label{section4A}
	\textbf{When the control direction is known, we add an additional adaptive law to estimate the time-varying control gain.} By the congelation of variables method and let $u=\hat\rho \bar u$,  we can rewrite $b(t)u$   as
	\begin{equation} b(t)u=\bar{u} +\Delta_b\hat{\rho}\bar{u}-\ell_b\left(\frac{1}{\ell_b}-\hat{\rho}\right)\bar{u}, 
	\end{equation}
	where   $\Delta_b=b(t)-\ell_b$, $\ell_b$ is an unknown constant, and  $\hat{\rho}$ is the estimate of $1/\ell_b$. Based on this treatment, we state the following theorem.

	\begin{theorem}\label{theorem1}
		Consider the parameter-varying strict-feedback system  (\ref{system}) with  unknown control gain magnitude yet known control direction. If Assumptions 1 and 2 are satisfied, then by using control law
		\begin{equation} \label{u4.1}
		\left\{\begin{array}{ll}
		\alpha_1(x_1,\hat{\theta}) =-(k_1+\zeta_1)z_1-w_1^{\top}\hat{\theta}\\
		\alpha_i(\underline{x}_i, \hat{\theta},\mu)=
		-(k_i+ \zeta_i) z_i -z_{i-1}-w_i^{\top}\hat{\theta}\\
		~~~~~~~~~~ ~ +\frac{\partial \alpha_{i-1}}{\partial\hat{\theta}}\Gamma \tau_i+ \sum_{j=2}^{i-1}\frac{\partial \alpha_{j-1}}{\partial \hat{\theta}}\Gamma \mu s_j w_i \\
		~~~~~~~~~~ ~   +\sum_{j=1}^{i-1}\frac{\partial \alpha_{i-1}}{\partial x_j}x_{j+1}+ \frac{\partial \alpha_{i-1}}{\partial \mu}\lambda \mu,\\ 
		\bar u= -  \kappa(\underline{x}_n,\hat{\theta},\mu) z_n \\
		u= \hat\rho \bar u
		\end{array}\right.
		\end{equation}
		where, for $i=1,\cdots,n-1$,  the auxiliary variables $z_i,~w_i,~\zeta_i$ and $\kappa$ are defined in (\ref{error variables})-(\ref{psi}), and the adaptive laws
		\begin{equation} \label{adaptive4.2} 
		\dot{\hat{\rho}}= -\gamma_{\rho}\operatorname{sgn}(\ell_b)\mu s_n \bar{u},  
		\end{equation}
		where the initial condition is chosen as $\hat{\rho}(0)>0~ \text{for} ~b(t)>0$ and $ \hat{\rho}(0)<0~ \text{for}  ~b(t)<0,$ and
				\begin{equation} \label{adaptive4.1}
			\dot{\hat{\theta}}=\Gamma \tau_n(\underline{x}_n,\hat{\theta},\mu),~~~  ~~\hat{\theta}(0)\geq 0, 
		\end{equation}
		then the equilibrium is globally exponentially stable, \textit{i.e.,} there exist  two positive numbers $N$ and $ \lambda$, such that $\|\underline{x}_n(t)\|\leq Ne^{- \lambda t}$; and all  signals in the system are ensured to be  bounded. Furthermore, $\lim_{t\rightarrow\infty}\hat{\theta}$ and $\lim_{t\rightarrow\infty}\hat{\rho}$ exist.
	\end{theorem}
	\textit{Proof:} The closed-loop dynamics of $s_i,~i=1,\cdots,n$ with the control law (\ref{u4.1}) are given by:
\begin{equation} \label{dynamics-si}
\begin{aligned}
\dot{s}_1=&-\left(k_1+\frac{n}{2}\delta_{\Delta_{\theta}}+\frac{1}{2\epsilon_{\psi}}+\frac{\delta_{\Delta_{\theta}}}{2}|W_1|_{_\mathsf{F}}^2\right)s_1+s_2\\
&+\mu w_1^{\top}(\ell_{\theta}-\hat{\theta})+\mu w_1^{\top} \Delta_{\theta},\\
\dot{s}_i=&-\left(k_i+\frac{n+1-i}{2}\delta_{\Delta_{\theta}}+\frac{1}{2\epsilon_{\psi}}+\frac{\delta_{\Delta_{\theta}}}{2}|W_i|_{_\mathsf{F}}^2\right)s_i\\
&-s_{i-1}+s_{i+1}+\mu w_i^{\top}(\ell_{\theta}-\hat{\theta})+\mu\frac{\partial \alpha_{i-1}}{\partial \hat{\theta}}\left(\Gamma\textcolor{blue}{\tau_i}-\dot{\hat\theta}\right)\\
&+\mu\sum_{j=2}^{i-1}\frac{\partial \alpha_{i-1}}{\partial \hat{\theta}}\Gamma \mu s_j w_i+\mu w_i^{\top}\Delta_{\theta},\\
\dot{s}_n=&-\kappa s_n +\lambda s_n-\kappa\Delta_{b}\hat{\rho}s_n-\mu {\ell_b}\left(\frac{1}{\ell_b}-\hat{\rho}\right)\bar{u}+\mu w_n^{\top}\hat{\theta}\\
&+\mu w_n^{\top}(\ell_{\theta}-\hat{\theta})+\mu w_n^{\top}\Delta_{\theta}-\mu\sum_{j=1}^{n-1}\frac{\partial \alpha_{n-1}}{\partial x_j}x_{j+1}\\
&-\mu\frac{\partial \alpha_{n-1}}{\partial \hat{\theta}}\dot{\hat{\theta}} {- \frac{\partial \alpha_{n-1}}{\partial \mu}\lambda \mu}.
\end{aligned}
\end{equation}
	Choosing a positive definite, radially unbounded function as a   Lyapunov function candidate
	\begin{equation}\label{V1}
	V=\frac{1}{2}\underline{s}_n^{\top}\underline{s}_n+\frac{1}{2}(\ell_{\theta}-\hat{\theta})^{\top}\Gamma^{-1}(\ell_{\theta}-\hat{\theta})+\frac{|\ell_b|}{2\gamma_{\rho}}\left(\frac{1}{\ell_b}-\hat{\rho}\right)^2.
	\end{equation}
	The derivative of $ V$ is given by
	\begin{equation}\label{37}
\begin{aligned}
		\dot V=&-\sum_{i=1}^{n-1}\left(k_i +\frac{1}{2\epsilon_{\psi}}+\frac{\delta_{\Delta_{\theta}}}{2}(|W_i|_{_\mathsf{F}}^2+n+i-1)\right)s_i^2\\
		&+s_{n-1}s_n  +\sum_{i=1}^{n}\mu s_iw_i^{\top}\Delta_{\theta}-\mu s_n  \frac{\partial \alpha_{n-1}}{\partial \mu}\lambda \mu\\ 
	& -\frac{ |\ell_b| }{\gamma_{\rho}}\left(\frac{1}{\ell_b}-\hat{\rho}\right)\left(\gamma_\rho \operatorname{sgn}(\ell_b) \mu s_n\bar u+\dot{\hat{\rho}}\right)\\
	& +(\ell_{\theta}-\hat{\theta})^{\top}\Gamma^{-1}\left(\Gamma\sum_{i=1}^{n }\mu s_i w_i  -\dot{\hat{\theta}}\right)-(\kappa -\lambda) s_n^2 \\
	&-\kappa\Delta_{b}\hat{\rho}s_n^2 +\mu s_nw_n^{\top}\hat{\theta}   -\mu s_n\sum_{j=1}^{n-1}\frac{\partial \alpha_{n-1}}{\partial x_j}x_{j+1}\\
	&+\mu s_n\frac{\partial \alpha_{n-1}}{\partial \hat{\theta}}\Gamma \tau_{n}+\mu s_2\frac{\partial \alpha_{1}}{\partial \hat{\theta}}\left(\Gamma{\tau_2}-\dot{\hat\theta}\right) \\
	&+\mu s_3\frac{\partial \alpha_{2}}{\partial \hat{\theta}}\left(\Gamma{\tau_3}-\dot{\hat\theta}\right) +\cdots +\mu s_{n}\frac{\partial \alpha_{n-1}}{\partial  \hat{\theta}}\left(\Gamma{\tau_{n}}-\dot{\hat\theta}\right)\\
	&+\mu s_3 \frac{\partial \alpha_{1}}{\partial \hat{\theta}}\Gamma \mu s_2 w_3 +\cdots+\mu s_n \sum_{j=2}^{n-1}\frac{\partial \alpha_{j-1}}{\partial \hat{\theta}}\Gamma \mu s_j w_n.
	\end{aligned}
	\end{equation}
	Since $	\dot{\hat{\theta}}=\Gamma\sum_{i=1}^{n }\mu s_i w_i $ and $\dot{\hat{\rho}}= -\gamma_{\rho}\operatorname{sgn}(\ell_b)\mu s_n \bar{u}$, then (\ref{37}) can be continued as follows:
		\begin{equation}\label{38}
\begin{aligned}
	\dot{V} 
	=&-\sum_{i=1}^{n-1}\left(k_i +\frac{1}{2\epsilon_{\psi}}+\frac{\delta_{\Delta_{\theta}}}{2}(|W_i|_{_\mathsf{F}}^2+n+i-1)\right)s_i^2 \\
	& +\sum_{i=1}^{n}\mu s_iw_i^{\top}\Delta_{\theta} -\kappa s_n^2 +\lambda s_n^2-\kappa\Delta_{b}\hat{\rho}s_n^2\\
	& + \mu s_n \left(z_{n-1}+ w_n^{\top}\hat{\theta}   - \sum_{j=1}^{n-1}\frac{\partial \alpha_{n-1}}{\partial x_j}x_{j+1}\right.\\
	&\left.+ \frac{\partial \alpha_{n-1}}{\partial \hat{\theta}}\Gamma \tau_{n} -  \sum_{j=2}^{n-1}\frac{\partial \alpha_{j-1}}{\partial \hat{\theta}}\Gamma \mu s_j w_{n}\right).
	\end{aligned} 
	\end{equation}	
	Reading Remark \ref{remmark W} shows that we can use the equality $w_i=W_i\underline{z}_i$ to arrive at the following  inequality
	\begin{equation}\label{39}
	\mu s_iw_i^{\top}\Delta_{\theta}\leq \frac{\delta_{\Delta_{\theta}}}{2}\left(|W_i|_{_{\mathsf{F}}}^2+1\right)s_i^2+\frac{\delta_{\Delta_{\theta}}}{2}\underline {s}_{i-1}^{\top}\underline{s}_{i-1}
	\end{equation}
	which will be canceled by the nonlinear damping terms embedded in the control laws. 
In addition, we know that
	 \begin{equation}\label{40}
	 {\mu s_n\psi=\mu s_n \bar\psi\underline{z}_n\leq\frac{1}{2}\left( \frac{1}{\epsilon_{\psi}}+\epsilon_{\psi}|\bar{\psi}|^{2} \right)s_n^2+\frac{1}{2\epsilon_{\psi}}\underline{s}_{n-1}^{\top} \underline{s}_{n-1}}.
	 \end{equation} 
Inserting  (\ref{u4.2}), (\ref{39}), and (\ref{40})  into (\ref{38}), we obtain
	\begin{equation}\label{dV4.2}
	\begin{aligned}
	\dot V\leq&-\sum_{i=1}^{n }k_i s_i^2  -\kappa(\underline{x}_n,\hat{\theta},\mu) \Delta_{b}\hat{\rho}s_n^2 .
	\end{aligned}
	\end{equation}
	The last term on the right-hand side of (\ref{dV4.2}) is always negative, because $\dot{\hat{\rho}}=\operatorname{sgn}(\ell_b)\gamma_{\rho}\kappa s_n^2   $, and
	\begin{itemize}
		\item Case 1 ($b(t)>0$): $\dot{\hat{\rho}}\geq 0$ and $\Delta_{b}>0$, then $\Delta_{b}\hat\rho(t)\geq 0$ for all $\hat{\rho}(0)>0$;
		\item Case 2 ($b(t)<0$): $\dot{\hat{\rho}}\leq 0$ and $\Delta_{b}<0$, then $\Delta_{b}\hat\rho(t)\geq 0$ for all $\hat{\rho}(0)<0$;
	\end{itemize}
	where $\hat{\rho}(0)$ is a design parameter, which can be selected according to $\operatorname{sgn}(b(t))$. According to the above analysis,  it follows that $-\kappa \Delta_{b} \hat{\rho} s_n^2\leq 0$,   implying that $\dot V \leq 0.$  Now we have
	\begin{equation}\label{sn4.1}
	\|\underline{s}_n\|\leq \sqrt{\|\underline{s}_n(0)\|^2+\frac{\|\ell_\theta-\hat{\theta}(0)\|^2}{2\Gamma}+\frac{|\ell_b|}{2\gamma_{\rho}}\left(\frac{1}{\ell_b}-\hat{\rho}(0)\right)^2}.
	\end{equation}
	In addition, it follows from (\ref{error variables}) and (\ref{exponential scaling}) that
	there exists a smooth and bounded mapping $M_{i-1}$, with $M_{i-1}(0,\hat{\theta},\underline{s}_{i-1})=0$,  such that
	\begin{equation}\label{xi4.1}
	x_i=\mu^{-1}s_i+M_{i-1}(\underline{x}_{i-1},\hat{\theta},\underline{s}_{i-1})\underline{x}_{i-1}
	\end{equation}
	and from (\ref{xi4.1}) that the initial condition satisfies
	\begin{equation}\label{initial-condition}
	s_i(0)=x_i(0)-M_{i-1}(\underline{x}_{i-1}(0),\hat{\theta}(0),\mu(0))\underline{x}_{i-1}(0).
	\end{equation} 
	Recursively applying (\ref{initial-condition})  results in that there exists  a nonnegative-valued continuous function  $N(\underline{x}_n(0),\hat{\theta}(0),\lambda)$ such that
	\begin{equation}\label{xi4.2}
	\|\underline{x}_n\|\leq N(\underline{x}_n(0),\hat{\theta}(0),\lambda)e^{-\lambda t}.
	\end{equation}
	Therefore, the closed-loop system is globally exponentially  stable.
	
	Furthermore, it also follows from (\ref{dV4.2}) that  $\hat\theta\in\mathcal{L}_{\infty}$ and $\hat{\rho}\in\mathcal{L}_{\infty}$. Recall from (\ref{u4.1}) that $\alpha_1=-(k_1+\zeta_1)z_1-w_1^{\top}\hat\theta$. Since $z_1=x_1$ and $w_1=\phi_1$, we see that $\alpha_1$ is bounded and therefore $z_2=x_2-\alpha_1$ is also bounded. The boundedness of $\tau_1=\mu w_1s_1=\Phi_1 s_1^2$ is then  established via (\ref{sn4.1}). In addition,  careful examination of (\ref{taui}) and (\ref{adaptive4.1}) reveals  that the quantity $\mu$ always appears multiplied by $z_i$, so in any instance where the $\mu z_i$ appears in the tuning functions or virtual control functions, it also can be guaranteed such functions are bounded. For example, from (\ref{u4.1}) we have $\alpha_2=-\zeta_2z_2-z_1-w_2^{\top}\hat\theta-w_1^{\top}\Gamma\tau_2-(k_1+\zeta_1)x_2$, its boundedness can be guaranteed by the boundedness of $\underline{x}_2$, $\underline{z}_2$, $\underline{w}_2$, $\hat{\theta}$ and $\tau_2$. Continuing in the same fashion, we prove that $\alpha_i (i=3,\cdots,n-1)$ and $u(t)$ are bounded. Rewrite (\ref{adaptive4.2}) we have $\dot{\hat{\rho}}=\gamma_{\rho}\kappa s_n^2$. Since $\underline{s}_n\in\mathcal{L}_2$, then $\dot{\hat{\rho}}\in\mathcal{L}_1$, then by using the argument similar to Theorem 3.1 in \cite{Krstic1996}, it is concluded that $\hat{\rho}$ has a limit as $t\rightarrow\infty$, establishing the same for $\hat{\theta}$. This completes the proof. $\hfill\blacksquare$

	\begin{remark}
		It is  crucial  to ascertain that the benefits of the proposed adaptive exponential control as stated in Theorem \ref{theorem1} are not achieved at the price of unbounded input and/or unbounded updating rate.  In fact, the design is based upon the exponential scaling (\ref{exponential scaling}), followed by a stabilizing adaptive control design for the scaled system.    As shown in (\ref{u3.1}), the product $e^{\lambda t} x$  in the update law is the scaled state $s$, which is kept bounded by (\ref{dV3.1.1}). The controller and adaptive laws for high-order systems inherit  this feature (see (\ref{u4.1}) and (\ref{adaptive4.1})) and hence the boundedness of which is naturally guaranteed.  In addition, after careful  examination of (\ref{error variables})-(\ref{adaptive4.1}), one can find that  the control input, estimated parameter, and parameter updating rate are all bounded.
	\end{remark}
	
	\begin{remark}
	In the absence of non-vanishing uncertainties, the closed-loop signals $\{s_i\}_{i=1}^n$ converge to zero asymptotically, while causing system states $\{x_i\}_{i=1}^n$ converge  at least $e^{-\lambda t}$ exponentially fast to zero. Clearly, if we let $\lambda=0$, then the system (\ref{system}) is asymptotically  stable. In this case, Theorem 1 is equivalent to [\citenum{chenkaiwen}, Proposition 1]. More importantly, since the virtual errors decrease gradually  with time as the time-varying gains increase, the ``peaking" phenomenon occurring in traditional high-gain feedback  does not exist here.
	\end{remark}
	
	\subsection{Adaptive control for parameter-varying nonlinear systems with unknown control direction}\label{section4B}
	\textbf{Significant challenge occurs in asymptotic control design when the control gain is unknown and time-varying.} This is particularly true in the context of exponential stabilizing control.  The main difference between  the design method for time-varying control coefficient in this subsection and the one for time-invariant control coefficient in   \cite{yexudong-auto,Huang2018,Liuyanjun,huachangchun} is that the former requires that the independent variable of the Nussbaum function must always be non-negative, thereby bringing new difficulties to the  control design. The proposed adaptive controller with a Nussbaum dynamic gain is given in the following theorem.
	\begin{theorem}\label{theorem2}
		Consider the parameter-varying strict-feedback system  (\ref{system}) with  unknown control gain magnitude and unknown control direction.  If Assumptions 1 and 3 are satisfied, then by using control law
		\begin{equation} \label{u5.1}
		\left\{\begin{array}{ll}
		\bar u= \kappa(\underline{x}_n,\hat{\theta},\mu) z_n \\
		\dot {\xi}=\mu s_n \bar u=\kappa s_n^2\\
		u= \mathcal{N}(\xi) \bar u
		\end{array}\right.
		\end{equation}
		and the virtual control laws as given in (\ref{u4.1}) and the adaptive law  as given in (\ref{adaptive4.1}),
		then the equilibrium is globally exponentially stable, \textit{i.e.,} there exist  two positive numbers $N$ and $\lambda$, such that $\|\underline{x}_n(t)\|\leq Ne^{-\lambda t}$; and all  signals in the system are ensured to be  bounded. Furthermore, $\lim_{t\rightarrow\infty}\hat{\theta}$ and $\lim_{t\rightarrow\infty}\hat{\rho}$ exist.
	\end{theorem}
	\textit{Proof:}	
	Recalling that
	\begin{equation}\label{46}
	\begin{aligned}
	u&=\mathcal{N}(\xi)\bar u\\
	\dot{\xi}&=\mu s_n \bar u.
	\end{aligned}
	\end{equation}	
The dynamic of $s_n$ under (\ref{46}) is given by
	 \begin{equation} 
	 	\begin{aligned}
	 	\dot{s}_n=&\mu \dot{z}_n+\dot{\mu}z_n\\
	 	=& \mu\left(b(t)u+\phi_n^{\top}\theta(t)-\sum_{j=1}^{n-1}\frac{\partial \alpha_{n-1}}{\partial x_j}\left(x_{j+1}+\phi_{j}^{\top}\theta(t)\right)\right.\\
	 	&\left.-\frac{\partial \alpha_{n-1}}{\partial \hat{\theta}}\dot{\hat{\theta}}-\frac{\partial \alpha_{n-1}}{\partial \mu}\lambda\mu
	 \right)+\lambda s_n.
	 	\end{aligned} 
	 \end{equation}
Since 
	 \begin{equation} 
	\mu s_n b(t)u=\mu s_n b(t)\mathcal{N}(\xi)\bar{u}=(b(t)\mathcal{N}(\xi)+1)\dot{\xi}-\mu s_n\bar u 
	\end{equation}	
then the following equation holds
	\begin{equation} \begin{aligned}
	s_n\dot{s}_n=&\left(b(t)\mathcal{N}(\xi)+1\right)\dot{\xi}+\mu s_n\left(-\bar u+\lambda z_n+w_n^{\top}\hat{\theta}\right.\\
	&\left.-\sum_{i=1}^{n-1}\frac{\partial\alpha_{n-1}}{\partial x_i}x_{i+1}-\frac{\partial\alpha_{n-1}}{\partial \hat{\theta}}\dot{\hat{\theta}}-\frac{\partial \alpha_{n-1}}{\partial \mu}\lambda\mu\right)\\
	&+\mu s_nw_n^{\top}(\ell_{\theta}-\hat{\theta})+\mu s_n w_n^{\top} \Delta_{\theta}.
	\end{aligned} 
	\end{equation}	
	Note that the dynamics of $s_i,~i=1,\cdots,n-1$ remain the same as in (\ref{dynamics-si}). 
	Choosing the   Lyapunov function candidate as
	\begin{equation}\label{V2}
	V=\frac{1}{2}\underline{s}_n^{\top}\underline{s}_n+\frac{1}{2}(\ell_{\theta}-\hat{\theta})^{\top}\Gamma^{-1}(\ell_{\theta}-\hat{\theta}).
	\end{equation}
	According to the proof of Theorem 1 and by virtue of the control law and adaptive law as shown in Theorem 2, we obtain
	\begin{equation}\label{dV5.11}
	\begin{aligned}
	\dot V\leq&-\sum_{i=1}^{n-1 }k_i s_i^2 -\frac{\delta_{\Delta_{\theta}}}{2}s_{n-1}^{\top} s_{n-1}-\frac{1}{2\epsilon_{\psi}}s_{n-1}^{\top}s_{n-1} \\
	&+\mu s_n\left(\lambda z_n-\bar u\right)+\mu s_n\psi  +\mu s_nw_n^{\top} \Delta_{\theta}\\
	&+\left(b(t)\mathcal{N}(\xi)+1\right)\dot{\xi} .
	\end{aligned}
	\end{equation}
	By recalling (\ref{exponential scaling}), (\ref{new regressor vectors}), (\ref{u4.2}),   (\ref{psi}) and (\ref{u5.1}), we have
	\begin{equation}
	\begin{aligned}
	&-\mu s_n \bar u+\mu s_n\psi  +\mu s_nw_n^{\top} \Delta_{\theta}\\
	&~~~~~~~~~~~~~\leq  -k _n  s_n^2  +  \frac{\delta_{\Delta_{\theta}}}{2}s_{n-1}^{\top} s_{n-1}+\frac{1}{2\epsilon_{\psi}}s_{n-1}^{\top}s_{n-1}
	\end{aligned}
	\end{equation}
	then (\ref{dV5.11}) becomes
	\begin{equation}\label{dV5.1}
	\begin{aligned}
	\dot V \leq \left(b(t)\mathcal{N}(\xi)+1\right)\dot{\xi}.
	\end{aligned}
	\end{equation}
	Note that $\dot\xi=\kappa s_n^2$, and we deliberately design $\kappa$ as $\kappa= k_n+\lambda+\frac{1}{2}\left( \delta_{\Delta_{\theta}}(|W_n|_{_{\mathsf{F}}}^2+1)+\frac{1}{\epsilon_{\psi}}+\epsilon_{\psi}|\bar{\psi}|^{2}\right) $ as shown in (\ref{u4.2}), therefore one can conclude that $\dot\xi=\kappa s_n^2 \geq 0$. By Lemma \ref{lemma1}, it follows   that $V(t)$ and $\xi(t)$ are bounded over $[0,\infty)$. The boundedness of $V(t)$ leads to the boundedness of $\underline{s}_n$ and $\hat{\theta}$. Note that the virtual control laws $\alpha_i$ and $\bar u$ and adaptive law $\dot{\hat{\theta}}$ in Theorem 2 have exactly the same form as that of Theorem 1, hence the boundedness of these signals can be can be directly derived from the proof of Theorem \ref{theorem1}. It is also follows that $\hat{\rho}$ has a limit as $t\rightarrow\infty$.
	The boundedness of $\xi(t)$ yields the boundedness of $\mathcal{N}(\xi)$, which further proves the boundedness of $u(t)$.
	In addition, from the proofs of Theorem \ref{theorem1}, we know that there exists a
	nonnegative-valued continuous function $N(\underline{x}_n(0),\hat{\theta}(0),\lambda)$ such that
	\begin{equation}\label{xi5.1}
	\|\underline{x}_n\|\leq N(\underline{x}_n(0),\hat{\theta}(0),\lambda)e^{-\lambda t}.
	\end{equation}
	Therefore, the closed-loop system is globally exponentially  stable. This completes the proof. $\hfill\blacksquare$
	
	
	\begin{remark}
		In \cite{chenkaiwen}, the original time-invariant parameter estimation via a certainty equivalence controller was extended to time-varying case, which shows that the system satisfying Assumptions 1-2 is asymptotically stable.  However, it should be noted that the asymptotic results in \cite{chenkaiwen}  are not attractive enough for some practical applications (such as high-performance robots\cite{sci-bird,sci-walk-robot}), because these applications often require the system to have a rapid transient response. To study the stabilizing control of systems to achieving rapid transient response, Theorem \ref{theorem1} proposes an exponentially stable controller for parameter-varying nonlinear systems, and Theorem \ref{theorem2} extends Theorem \ref{theorem1} to parameter-varying nonlinear systems with  unknown control directions.
	\end{remark}

	\begin{remark}
	Unlike prescribed performance control (PPC) methods, in which, the system output is guaranteed to evolve within a performance function and ultimately decay to  a residual set, where the control parameters are determined according to system initial condition, our method ensures that all system states converge to zero within an exponential decay  rate. On the other hand, a common technology adopted in PPC is to transform the  ``constrained" system into an equivalent ``unconstrained" one via a coordinate transformation; however, the key idea in this article is scaling the virtual control errors by a time-varying function and the stability analysis is based on a time-varying Lyapunov function. Furthermore, we establish the global stability of the closed-loop system, without requiring an \textit{a priori} knowledge of the initial condition.
	\end{remark}

	\section{Simulations}\label{section5}
	Consider the scenario in which a high-performance airplane flying at high angle of attack aims at stabilizing its wing rock unstable motion.  A single degree of freedom model is extracted from   \cite{Krstic1996model}, as follows
	\begin{equation}\label{wink}
	\begin{aligned}
	&\dot \phi=p\\
	&\dot p= \frac{\bar q S b}{I_x}\left(0.5C_{l_1}\phi\sin(\alpha)+\frac{C_{l_2}pb}{2V}+C_{\delta_A}\right)
	\end{aligned}
	\end{equation}
	where $\alpha$ is angle of attack in degrees, $\phi$ is the roll angle in radians, and $p$ is the roll rate in radians per second. The constants $\bar q, S, b, I_x$ and $V$ are the dynamic pressure, wing reference area, wing span, roll moment of inertia, and freestream air speed, respectively. The coefficients $C_{l_1}$ and $C_{l_2}$ are the rolling moment   derivatives, $C_{\delta_A}$ is the control surface.
	
	The parametric strict-feedback form of the wing rock model (\ref{wink}) by letting $x_1=\phi$, $x_2=p$ and $C_{\delta_A}=u$ is
	\begin{equation}
	\begin{aligned}
	\dot x_1&=x_2+\phi_1^{\top}\theta(t)\\
	\dot x_2&=b(t)u+\phi_2^{\top}\theta(t)
	\end{aligned}
	\end{equation}
	where $\phi_1=0$, $\phi_2=[x_1,x_2]^{\top}$, $b(t)= {\bar q S b}/{I_x}$,  and $\theta(t)=[\theta_1(t),\theta_2(t)]^{\top}=[0.5C_{l_1}\sin(\alpha)\bar{q}Sb/I_x,C_{l_2}\bar q Sb^2/(2I_x V)]^{\top}$. Note that \cite{Krstic1996model} provides the following wind-tunnel data at angle of attack of $\alpha=30^{o}$: $\theta_1=-26.6667$ and $\theta_2=0.67485$. Taking into account that the change of the attack angle will cause $\theta$ to change, therefore we assume in the simulation that $\theta_1$ and $\theta_2$ will periodically change by $\pm 2\%$ on the basis of the experimental data\footnote{{Here $b(t)$ and $\theta(t)$ are fast time-varying parameters since they are only piecewise continuous and may undergo sudden changes. Therefore, most adaptive schemes  (see, for instance, \cite{Goodwin-nonPE,Marino-TAC,dongxiwang,song1992tac,Zhao}) are not available because those methods require the  parameters be constant or slow time-varying.}}, \textit{i.e.,} $\theta_i(t)=\theta_i+2\%\theta_i\operatorname{sgn}(\sin(3t))$ for $i=1,2.$ In addition, except only knowing that $b(t)\neq 0$, we have not obtained other information about $b(t)$ from the experimental data. In other words, the control direction and the control magnitude  are unknown. Here  we set the control coefficient $b(t)$ as  $b(t)=-2+0.2\operatorname{sgn}(\sin(3t))\cos(t)$ to simulate the parameter changes at different angles of attack. 
 Note that the parameters $\theta(t)$ and $b(t)$ comprise of a constant nominal part and a time-varying part designed to destabilize the system. 
	For the system under consideration, it is readily verified that Assumptions 1-3  are satisfied, thus the control scheme proposed in Theorems 1-2 can be directly applied to stabilize (\ref{wink})  exponentially. Here we consider three different controllers:
	\begin{itemize}
			\item \textbf{Controller 1}: the adaptive asymptotic controller (AC) proposed in \cite{chenkaiwen}; 
		\item \textbf{Controller 2}: the adaptive controller with  exponential convergence rate proposed in Theorem 1; 
		\item \textbf{Controller 3}: the adaptive Nussbaum controller with  exponential convergence rate proposed in Theorem 2. 
	\end{itemize}  
 For fair comparison, we set   $\delta_{\Delta_{\theta}}=0.6$, $k_1=k_2=1$, $[\hat\theta_1(0);\hat\theta_2(0)]=[0;0]$,   $\lambda=0.6$, $\Gamma=0.001I$ and $[x_1(0);x_2(0)]= [-1;2.5]$ for all Controllers.  In addition, for Controllers 1 and 2, we set $\hat{\rho}(0)=-0.3$,  and for Controller 3, we set $\xi(0)=0$ and choose  the enhanced Nussbaum function  as $\mathcal{N}(\xi)=\sin(\xi)\exp ({\xi^2})$ according to the Example 5.3 in \cite{chenzhiyongauto}.
	\begin{figure} [!]
		\begin{center}
			\includegraphics[height=7cm]{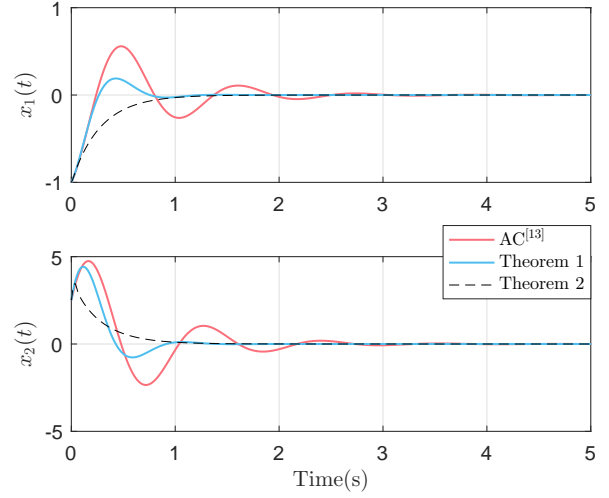}
			\caption{The responses of $x_1(t)$ and $x_2(t)$ under different Controllers.}
		\end{center}\label{fig1}
	\end{figure}
	\begin{figure} [!]
		\begin{center}
			\includegraphics[height=3.85cm]{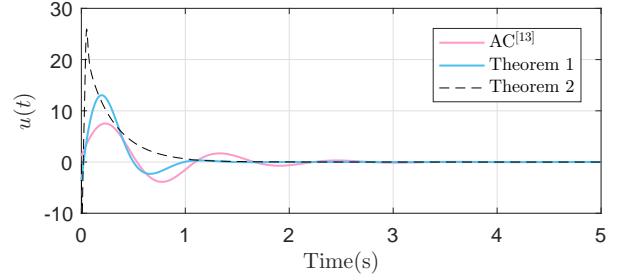}
			\caption{The  response  of $u(t)$ under different  different Controllers.}
		\end{center}\label{fig2}
	\end{figure}
	\begin{figure} [!]
		\begin{center}
			\includegraphics[height=7cm]{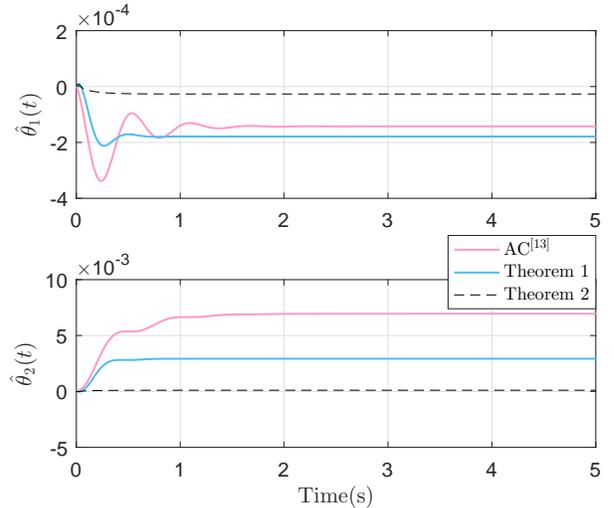}
			\caption{The  responses  of $\hat{\theta}_1(t)$ and $\hat{\theta}_2(t)$ under  different Controllers.}
		\end{center}\label{fig3}
	\end{figure}
	\begin{figure} [!]
		\begin{center}
			\includegraphics[height=7 cm]{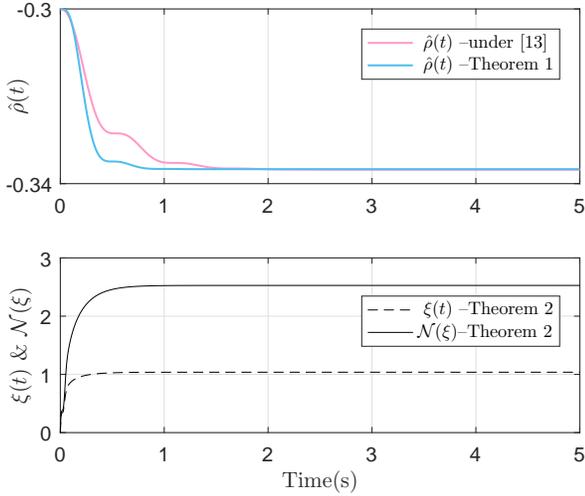}
			\caption{The responses of controller parameters $\hat{\rho}(t),~\xi(t)$ and $\mathcal{N}(\xi)$.}
		\end{center}\label{fig4}
	\end{figure}
	\begin{figure} [!]
		\begin{center}
			\includegraphics[height=7cm]{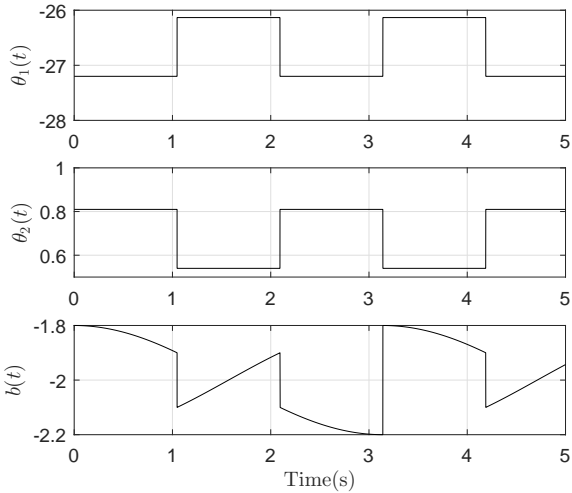}
			\caption{The  responses  of system parameters $\theta_1(t),~\theta_2(t)$ and $b(t)$.}
		\end{center}\label{fig5}
	\end{figure}

	The  evolutions of the system states and control input are illustrated in Figs 1-2, respectively. The evolution of adaptive parameters $\hat{\theta}_1(t)$ and $\hat\theta_2(t)$ are illustrated in Fig. 3; and the evolutions of $\hat{\rho}(t)$, $\xi(t)$ and $\mathcal{N}(\xi)$ are illustrated in Fig. 4. In addition, the time-varying system parameters $\theta_1(t)$, $\theta_2(t)$ and $b(t)$ are illustrated in Fig 6.  From simulation results, one can find that $1)$ all signals are bounded and the independent variable of the Nussbaum function is always non-negative; $2)$ under Theorems 1-2, the system state converges to zero at an exponential speed, while under \cite{chenkaiwen}, the system state converges to zero at a relatively slow speed, and the overshoot is larger than the former;  and $3)$ the ``peaking" phenomenon  does not appear in Controller 1, but it appears in Controller 3.  As a matter of fact, how to weaken the ``peaking" phenomenon produced by Nussbaum-gain technology (Controller 3) is a challenging yet meaningful problem.  In short, the above results illustrate the superiority and effectiveness of our approaches.

	\section{conclusion}\label{section6}
	The notion of accelerating convergence process by making use of rate function transformation and the technique of handling time-varying parameters via congelation of variables method are quite  appealing in developing accelerated control for parameter-varying strict-feedback systems, which, together with the integration of enhanced Nussbaum function, could allow new adaptive control (as simple as the traditional adaptive control) to be developed for a class of nonlinear systems with unknown time-varying parameters in feedback path and input path, yet involving time-varying  control gain that is unknown in sign and in magnitude. The stability conditions have been verified with the help of suitable time-varying Lyapunov functions. Future work includes seeking some suitable ways to guarantee the parameters converge exponentially to their desired values (see \cite{2022-Glushchenko}) and/or reduce the waste of computing resource caused by continuous-time scheme.
	
	


\end{document}